\documentclass[UKenglish,cleveref, autoref, thm-restate, algorithmic, algorithm]{lipics-v2021}
\usepackage[ruled, noend, algo2e]{algorithm2e}
\usepackage[noend]{algorithmic} 

\usepackage{amsmath, amsfonts, amssymb}
\usepackage{todonotes}
\usepackage{booktabs}
\usepackage{numprint}
\usepackage{changepage}
\usepackage{tcolorbox}
\usepackage{comment}
\usepackage{siunitx}
\newcommand{\cpp}{\textsf{C}\texttt{++}\xspace}
\newcommand{\eps}{\varepsilon}

\newcommand{\algorithmCCHHQRS}{\textsf{Fractional}\xspace}
\newcommand{\algorithmPCCHHQRS}{\textsf{PackedFractional}\xspace}
\newcommand{\algorithmPCCHHQRSList}{\textsf{PackedFractional, list}\xspace}
\newcommand{\competitorImprovedBFS}{\textsf{ImprovedDynOpt}\xspace}
\newcommand{\competitorStrongBFS}{\textsf{StrongDynOpt}\xspace}
\newcommand{\competitorLimitedBFS}{\textsf{BFS20}\xspace}

\title{From Theory to Practice: Engineering Approximation Algorithms for Dynamic Orientation} 
\titlerunning{Engineering Approximation Algorithms for Dynamic Orientation}
\supplement{}

\supplementdetails[subcategory={Source Code}, cite={}, swhid={}]{Software}{https://github.com/DynGraphLab/DynDeltaApprox}
\supplementdetails[subcategory={Dataset}]{Data Set}{https://zenodo.org/records/15785840}

\author{Ernestine Grossmann}{Heidelberg University, Germany}{e.grossmann@informatik.uni-heidelberg.de}{https://orcid.org/0000-0002-9678-0253}{}

\author{Henrik Reinstädtler}{Heidelberg University, Germany}{henrik.reinstaedtler@informatik.uni-heidelberg.de}{https://orcid.org/0009-0003-4245-0966}{}

\author{Eva Rotenberg}{IT University of Copenhagen, Denmark}{erot@itu.dk}{https://orcid.org/0000-0001-5853-7909}{}

\author{Christian Schulz}{Heidelberg University, Germany}{christian.schulz@informatik.uni-heidelberg.de}{https://orcid.org/0000-0002-2823-3506}{}

\author{Ivor van der Hoog}{IT University of Copenhagen, Denmark}{ivva@itu.dk}{https://orcid.org/0009-0006-2624-0231}{This project has received funding from the European Union's Horizon 2020 research and innovation programme under the Marie Sk\l{}odowska-Curie grant agreement No 899987.}

\author{Juliette Vlieghe}{IT University of Copenhagen, Denmark}{jmvvl@dtu.dk}{https://orcid.org/0009-0004-0079-8523}{}

\funding{{\it Ivor van der Hoog}, {\it Eva Rotenberg}, and {\it Juliette Vlieghe} are grateful to the VILLUM Foundation for supporting this research via Eva Rotenberg's Young Investigator grant (VIL37507) ``Efficient Recomputations for Changeful Problems''. Part of this work took place while these authors were affiliated with the Technical University of Denmark.
\\
{\it Ernestine Grossmann}, {\it Henrik Renstädtler}, and {\it Christian Schulz} acknowledge support by DFG grant \hbox{SCHU 2567/8-1}.
}

\authorrunning{E. Grossmann, H. Reinstädtler, E. Rotenberg, C. Schulz, I. van der Hoog and J. Vlieghe} 

\Copyright{Ernestine Grossmann, Henrik Reinstädtler, Eva Rotenberg, Christian Schulz, Ivor van der Hoog and Juliette Vlieghe} 



\category{} 
\hideLIPIcs
\nolinenumbers

\EventEditors{Anne Benoit, Haim Kaplan, Sebastian Wild, and Grzegorz Herman}
\EventNoEds{4}
\EventLongTitle{33rd Annual European Symposium on Algorithms (ESA 2025)}
\EventShortTitle{ESA 2025}
\EventAcronym{ESA}
\EventYear{2025}
\EventDate{September 15--17, 2025}
\EventLocation{Warsaw, Poland}
\EventLogo{}
\SeriesVolume{351}
\ArticleNo{63}

\begin{document}

\maketitle

\begin{abstract}
Dynamic graph algorithms have seen significant theoretical advancements, but practical evaluations often lag behind. This work bridges the gap between theory and practice by engineering and empirically evaluating recently developed approximation algorithms for dynamically maintaining graph orientations. 
We comprehensively describe the underlying data structures, including efficient bucketing techniques and round-robin updates. 
Our implementation has a natural parameter $\lambda$, which allows for a trade-off between algorithmic efficiency and the quality of the solution. In the extensive experimental evaluation, we demonstrate that our implementation offers a considerable speedup. Using different quality metrics, we show that our implementations are very competitive and can outperform previous methods.
Overall, our approach solves more instances than other methods while being up to 112 times faster on instances that are solvable by all methods compared.
\end{abstract}
\clearpage

\section{Introduction}

A (fully) dynamic graph algorithm is a data structure designed to support edge insertions, edge deletions, and answer specific queries. 
While there has been extensive research on dynamic graph problems, many of these theoretically efficient algorithms have not been implemented or empirically evaluated.
In this work, we  dynamically maintain an orientation of a graph. Let $G =(V, E)$ be an undirected graph with $n$ vertices and $m$ edges. 
A \emph{fractional orientation} $\overrightarrow{G}$ of $G$ assigns to each edge $\{ u, v \} \in E$ two non-negative weights $d(u \rightarrow v)$ and $d(v \rightarrow u)$ such that $d(u \rightarrow v) + d(v \rightarrow u) = 1$.
The \emph{fractional out-degree} $d^+(u)$ of a vertex $u$ is the sum over all edges $\{ u, v \} \in E$  of $d(u \rightarrow v)$.
An \emph{integral orientation} is a fractional orientation in which all weights are either $0$ or $1$. We will denote by $d^*(u)$ the out-degree of a vertex $u$ if the orientation is integral. 
Graph orientations have a wide range of applications:

\subparagraph{Fractional Orientations and Graph Mining.}
Given a graph $G$, a \emph{density decomposition} of~$G$ is a nested set of subgraphs of~$G$ of decreasing density. 
Finding dense subgraphs lies at the core of graph mining~\cite{gionis2015dense}.
Density decompositions were first applied in finding clusters in web-data~\cite{gibson1998inferring, gibson2005discovering, kleinberg1998authoritative,kleinberg1999web} but quickly embedded themselves in pattern mining techniques of other fields such as social graph mining~\cite{angel2014dense, chen2010dense, dourisboure2007extraction, kumar2006structure, sozio2010community},
bioinformatics~\cite{bader2003automated, fratkin2006motifcut, hagmann2008mapping, saha2010dense, zhang2005general}, finance~\cite{du2009migration} and even drawing algorithms for large graphs~\cite{alvarez2005large}.
Danisch, Chan, and Sozio~\cite{Prac2} propose a state-of-the-art density composition of a graph, which they call the \emph{local density}. 
Due to its applicability, there exist many \emph{static} implementations to efficiently approximate (variants of) this decomposition~\cite{harb2022faster, ma2020efficient, sun2020kclist++, xu2023efficient}.
There exists a 1:1 correspondence between this density decomposition and fractional orientations. 
Indeed, consider a fractional orientation $\overrightarrow{G}$ that minimizes the sum-of-squares metric defined by $\texttt{sum}^2(\overrightarrow{G}) := \sum_{v \in V} \left( d^+(v) \right)^2$.
The decomposition by Danish, Chan, and Sozio is the partition of vertices by their out-degree.

\subparagraph{Integral Orientations and Dynamic Efficiency.}
A crucial {building block} for efficient, fully dynamic algorithms is efficient, dynamic edge-orientations.
Given an undirected graph $G$, the \emph{edge orientation problem} asks for an orientation of $G$ in which the maximum out-degree of the vertices is minimized.
Many real-world graphs are relatively sparse and allow for edge orientations where the maximum out-degree is considerably smaller than the maximum degree of the graph. 
Low out-degree integral orientations allow for efficient and low-memory adjacency queries. This, in turn, has many applications, including dynamic map labeling, graph colouring, maximal matchings, and finding $k$-cliques~\cite{dhulipala2021parallel}.
Other example applications include Neiman and Solomon \cite{DBLP:journals/talg/NeimanS16} who have shown how to maintain a maximal matching in $O(\log n / \log \log n)$ amortized time using a dynamic edge orientation algorithm, dynamic matrix vector multiplication~\cite{DBLP:conf/icalp/KopelowitzKPS14}, or Kowalik and Kurowski~\cite{DBLP:journals/talg/KowalikK06} who use dynamic edge orientations to answer shortest-path queries of length at most $k$ in $O(k)$ time in planar graphs. 
The aforementioned algorithms use time proportional to the maximum out-degree in the orientation. Thus, it becomes important to minimize $\texttt{max}(\overrightarrow{G}) := \max_{v \in V} d^+(v)$.

\subparagraph{Minimax versus Least Square Regression.}  In graph mining, we are interested in orientations $\overrightarrow{G}$ that minimize  $\texttt{sum}^2(\overrightarrow{G})$.
Minimising this measure optimises for the ``average'' out-degree which identifies the density of the neighborhood of each vertex. 
On the other hand, many dynamic algorithmic application domains are interested in orientations $\overrightarrow{G}$ that minimize  $\max(\overrightarrow{G})$. 
Danisch, Chan, and Sozio prove that, in this case, the least square regression is strictly more stringent than the minimax regression as any fractional orientation $\overrightarrow{G}$ minimizing $\texttt{sum}^2(\overrightarrow{G})$ also minimizes $\max(\overrightarrow{G})$~\cite{Prac2}. 
 {The value of $\max(\overrightarrow{G})$ equals the pseudoarboricity of the graph. The (pseudo-)arboricity is defined as the number of (pseudo-)forests that the graph can be partitioned into. The arboricity is either equal to  the pseudoarboricity or plus one~\cite{picard1982network}.  }

\subparagraph{Prior Work in Theory. }
Theoretical dynamic algorithms typically maintain a fractional orientation $\overrightarrow{G}$ over $G$.
A naive rounding scheme then converts each fractional orientation $\overrightarrow{G}$ to an integral orientation $\overrightarrow{G}^*$ where for all $v \in V$, the out-degree $d^*(v)$ is at most $2 d^+(v)$. 
 {A simple rounding scheme rounds all fractional edges with a value greater than $\frac{1}{2}$ and ones with equal value by a deterministic scheme, like the assigned vertex id.  }
Christiansen and Rotenberg~\cite{christiansenICALP} propose a dynamic polylog-time rounding algorithm where for all $v \in V$, $d^*(v) \leq d^+(v) + 1$.
The state-of-the-art dynamic theoretical algorithm to maintain a fractional orientation that minimizes $\texttt{max}(\overrightarrow{G})$ is by Chekuri et al.~\cite{chekuri2024adaptive}.
They dynamically maintain a fractional orientation $\overrightarrow{G}$ where for all $\overrightarrow{G}'$, $\texttt{max}(\overrightarrow{G}) \leq (1+\eps) \cdot \texttt{max}(\overrightarrow{G}')$.
Their algorithm uses linear space and has $O( \eps^{-6} \log^4 n)$ update time. 
Christiansen, van der Hoog, and Rotenberg~\cite{christiansen2025local} show that the algorithm from~\cite{chekuri2024adaptive}, in actuality, maintains a fractional orientation $\overrightarrow{G}$ such that for all $\overrightarrow{G}'$, $\texttt{sum}^2(\overrightarrow{G}) \leq (1+\eps) \cdot \texttt{sum}^2 (\overrightarrow{G}')$. Thus, the state-of-the-art theoretical algorithm maintains a fractional orientation that $(1+\eps)$-approximates the optimal solution for both our metrics of interest in $O( \eps^{-6} \log^4 n)$ time per update. 

\subparagraph{Prior Work in Practice.}
In contrast to theoretical algorithms, most practical methods for computing edge-orientations on dynamic graphs maintain an integral orientation. We now discuss some experimentally evaluated approaches.
Brodal and Fagerberg~\cite{dblp:conf/wads/brodalf99} were the first to consider the problem in the dynamic case. With a bound $c$ on the arboricity of the graph as input, their approach
supports adjacency queries in time $O(c)$, edge insertions
in amortized time $O(1)$, and edge deletions in
amortized time $O(c + \log n)$ time. 

Berglin and Brodal~\cite{DBLP:journals/algorithmica/BerglinB20} allow a \emph{worst-case} user-specific trade-off between out-degree and running time of the operations. Their algorithm maintains an $O(\alpha+\log n)$ orientation in $O(\log n)$ worst-case time or an $O(\alpha \log^2 n)$ orientation in constant worst-case time.
Borowitz et al.~\cite{dblp:conf/acda/borowitzg023} experimentally evaluated different heuristics and approximation algorithms.
This evaluation includes new approaches presented by the authors, which are based on a BFS strategy as well as the algorithm by Brodal and Fagerberg~\cite{dblp:conf/wads/brodalf99} and the $K$-Flip algorithm introduced by Berglin and Brodal \cite{DBLP:journals/algorithmica/BerglinB20}. In this experimental evaluation, the BFS approach performed best.
Recently, Gro{\ss}mann et al.~\cite{grossmann2025engineering} presented a dynamic algorithm to solve the problem exactly. The worst-case complexity of this method for insertions is $O(m)$, while for deletion it is amortized $O(m)$. Their experiments show that their exact algorithm achieves up to 32\,\% lower running times than the inexact BFS approach by Borowitz et al.~\cite{dblp:conf/acda/borowitzg023}.

\subparagraph{Contribution.}
We engineer an implementation of the recently proposed approximation algorithms by Chekuri et al.~\cite{chekuri2024adaptive}. In Section~\ref{sub:chekuri}, we provide crucial details required for an implementation.
A straightforward implementation of the algorithms would be quite slow and memory-intensive. This is because~\cite{chekuri2024adaptive} defines a large  (constant) number of values and pointers to  guarantee that all algorithmic operations take theoretically $O(1)$ time. Real-world graphs  contain many edges compared to the number of available bits in memory. Storing up to eight integers and pointers per edge is infeasible and a clear bottleneck.

Thus, in Section~\ref{sec:improvement}, we present engineering aspects to adapt the algorithm from~\cite{chekuri2024adaptive} to reduce space use and considerably improve its efficiency. 
As a consequence, we obtain two implementations of~\cite{chekuri2024adaptive} to maintain a fractional orientation $\overrightarrow{G}$ of a dynamic graph $G$ that simultaneously approximately minimizes both $\texttt{sum}^2 (\overrightarrow{G})$ and $\texttt{max} (\overrightarrow{G})$. 
We note that these existing algorithms maintain an integral orientation $\overrightarrow{G}^*$ and are designed to minimize $\texttt{max} (\overrightarrow{G}^*)$.
To provide a fair comparison between our implementation and the current state-of-the-art, we convert our fractional orientation $\overrightarrow{G}$ to an integral orientation through the naive rounding scheme. 
Experiments in Section~\ref{sec:experiments} show that our implementations offer a considerable speedup of up to 112.17 times whilst being very competitive on the measure $\texttt{max} (\overrightarrow{G}^*)$.
Our algorithm performs considerably better than previous implementations when using $\texttt{sum}^2(\overrightarrow{G}^*)$ as the quality metric. 
Our implementation has a natural parameter $\lambda$, which allows for a trade-off between algorithmic efficiency and the quality of the solution.

\section{Preliminaries}\label{sec:prelim}

Let $G = (V, E)$ be a dynamic, unweighted, undirected graph subject to 
edge insertion and edge deletion. 
We say that $|V| = n$ at the time of an update. 
A \emph{fractional} orientation $\overrightarrow{G}$ of $G$ assigns for every edge $\{ u, v \}$ two non-negative weights: $d(u \to v)$ and $d(v \to u)$ such that $d(u \to v) + d(v \to u) = 1$. 
The \emph{out-degree} of a vertex $u$ is defined as $d^+(u) = \sum_{v \in V} d(u \to v)$. 
An \emph{integral orientation} is a fractional orientation where for all $\{ u, v \} \in E$, $d(u \to v) \in \{0, 1 \}$. 
To distinguish between integral and fractional orientations, we denote by $d^*(u)$ the out-degree of a vertex $u$ if the orientation is integral. 
In integral orientations, we reference directed edges as $\overrightarrow{uv}$ and $\overrightarrow{vu}$. 
A naive rounding scheme converts each fractional orientation $\overrightarrow{G}$ into an integral orientation $\overrightarrow{G}^*$ where for all $v \in V$, the out-degree $d^*(v)$ is at most $2 d^+(v)$.

\subparagraph{Quality Measures.}
We denote by $\mathbb{O}(G)$ the infinite-size set of all fractional orientations of $G$. For any orientation $\overrightarrow{G} \in \mathbb{O}(G)$ we define two quality measures: 
\[
\texttt{max}(\overrightarrow{G}) := \max_{v \in V} \, d^+(v), \hspace{1.7cm} \textnormal{and, } \hspace{1.5cm} 
\texttt{sum}^2(\overrightarrow{G}) := \sum_{v \in V} \, \left( d^+(v) \right)^2.
\]

\noindent
These measures induce respectively a linear and a convex optimization problem over $\mathbb{O}(G)$:
\[
\Delta^{\min}(G) := \min_{\overrightarrow{G} \in \mathbb{O}(G)} \texttt{max}(\overrightarrow{G}), \hspace{1cm} \textnormal{and, } \hspace{1cm} 
\Delta^{ local}(G) := \min_{\overrightarrow{G} \in \mathbb{O}(G)} \texttt{sum}^2(\overrightarrow{G}).
\]


\subparagraph*{Problem Statement.}
State-of-the-art edge orientation implementations use integral orientations.
We therefore maintain  for a dynamic unweighted graph $G$ either:
\begin{itemize}
    \item an {integral orientation} $\overrightarrow{G}^*$ that approximately minimizes $\texttt{max}(\overrightarrow{G}^*)$, or,
    \item  an integral orientation $\overrightarrow{G}^*$ that approximately minimizes $\texttt{sum}^2(\overrightarrow{G}^*)$
\end{itemize}
\noindent
To this end, we dynamically maintain a fractional orientation $\overrightarrow{G}$ and apply naive rounding.

\subparagraph{Fairness.}
Sawlani and Wang~\cite{sawlani2020near} say that $\overrightarrow{G} \in \mathbb{O}(G)$ is \emph{locally fair} whenever $d(u \rightarrow v) > 0$ implies that $d^+(u) \leq d^+(v)$. 
They prove that if $\overrightarrow{G}$ is locally fair, then $\texttt{max}(\overrightarrow{G}) = \Delta^{\min}(G)$. 
Christiansen, van der Hoog, and Rotenberg~\cite{christiansen2025local} show an even stronger correlation between fair orientations and the sum-of-squares measure as $\overrightarrow{G}$ is locally fair if and only if $\texttt{sum}^2(\overrightarrow{G}) = \Delta^{ local}(G)$.
A locally fair orientation cannot be maintained without requiring $\Omega(n)$ update time. Hence, several works relax the notion of fairness to approximate the local fairness and thereby $\Delta^{\min}(G)$ and $\Delta^{ local}(G)$~\cite{BeraBCG22, borradaile2017egalitarian, borradaile2018density, christiansen2025local, frank2023fair, sawlani2020near, zhang2024efficient}.
For a fixed  $\theta \in \{0, 1 \}$ and $\lambda \in [0, 1)$,
Chekuri et al.~\cite{chekuri2024adaptive} define $\overrightarrow{G}$ as $\lambda$-fair if:
$
d(u \rightarrow v) \Rightarrow d^+(u) \leq (1+\lambda) \cdot d^+(v) + \theta.$

\noindent
If we subsequently smartly choose $\lambda \in \Theta\left( \frac{\eps^2}{\log n}\right)$ and $\theta = 0$, then any $\lambda$-fair fractional orientation $\overrightarrow{G}$ guarantees that $\texttt{max}(\overrightarrow{G})$ is a $(1+\eps)$-approximation of $\Delta^{\min}(G)$~\cite{chekuri2024adaptive} and that 
$\texttt{sum}^2(\overrightarrow{G})$ is a $(1+\eps)$-approximation of $\Delta^{local}(G)$~\cite{christiansen2025local}. 
If $\theta = 1$, the $\lambda$-fair orientation may be integral instead, which is considerably easier to maintain. Setting $\theta = 1$ does come at a cost in quality, as the approximation becomes an additive error of $O(\log n)$.

\subsection{The Algorithm for Maintaining a $\lambda$-fair Orientation by~\cite{chekuri2024adaptive}}
\label{sub:chekuri}

\begin{figure}
    \centering
    \includegraphics[]{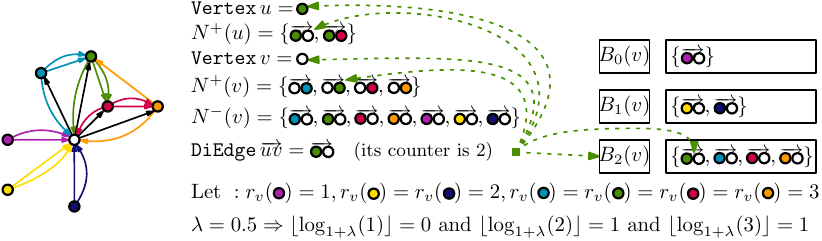}
    \caption{A graph $G_b$ for $b = 2$ and an orientation $G_b^*$. 
    We illustrate the data structure from~\cite{chekuri2024adaptive}.   
    Observe that the \texttt{DiEdge} $\overrightarrow{uv}$ stores six pointers in this data structure to allow constant-time updates.
    }
    \label{fig:data_types}
\end{figure}

Let $\lambda \in [0, 1]$ be some quality parameter that the user fixed beforehand.
Then, the algorithm by~Chekuri et al.~\cite{chekuri2024adaptive}, on a high level, inserts (or deletes) directed edges in $G_b^*$ (which is an integral orientation over the graph $G$ where each edge is duplicated $b$ times). They orient the edges in such way that they maintain an (integral) $\lambda$-fair orientation. 
After inserting a directed edge $\overrightarrow{uv}$, the out-degree of $u$ increases. So, there may exist some edge $\overrightarrow{ux}$ for which $d^*(u) > (1+\lambda)d^*(x)$.
The algorithm then flips this edge, restoring the out-degree of $x$, and recurses on the edge-insertion $\overrightarrow{xu}$. 
Instead of spending $O(d^*(u))$ time to inspect all out-edges $x$ of $u$, the authors show that it suffices to loop over a fraction of the out-edges of $u$ in round-robin fashion.
To make all supporting operations in this algorithm run in constant time, the algorithm uses elaborate data structures, data types, and information updates. 
We explain these first.

For $b \in \mathbb{N}$, denote by $G_b$ the graph $G$ where every edge is duplicated $b$ times.
Each edge insertion in $G$ thereby inserts $b$ edges into $G_b$.
We maintain an integral orientation $G_b^*$ over $G_b$ where $d^*_b(v)$ denotes the out-degree.
$G_b^*$ induces a fractional orientation over $G$ and we obtain an integral orientation ${G}^*$ of $G$ from $G_b^*$ by naively rounding each edge.

We denote for all $u \in V$ by $N^+(u)$ all vertices $v$ for which $G_b^*$ has at least one edge $\overrightarrow{uv}$ (see Figure~\ref{fig:data_types}).
The set $N^-(u)$ are all vertices $v$ where $\overrightarrow{G_b}$ includes at least one edge $\overrightarrow{vu}$. 
Note that $d^*_b(u)$ is not necessarily equal to $|N^+(u)|$ because $G_b$ is a multigraph.

 \subparagraph{Information Updates.}
 The update algorithm to $G_b^*$ implements a lazy information update scheme where adding an edge $\overrightarrow{uv}$ to $G_b^*$ increments a counter recording the number of edges $\overrightarrow{uv}$ that $G_b^*$ contains without updating all auxiliary information and data structures.

Formally we define an \emph{information update} (we forward reference Algorithm~\ref{alg:information}). 
The algorithm may invoke an information along a directed edge. 

 \begin{definition}
  Let $d^*_b(w)$ be the out-degree of $w$ at the time of the information update. 
 We define the \emph{log-degree} $r_u(w)$ as the floor of the base $(1+\lambda)$ logarithm of $d^*_b(w)$ \emph{at the time of an information update along} $\overrightarrow{wu}$. 
 Or formally: $
 r_u(w) = \lfloor \log_{1 + \lambda} d^*_b(w) \rfloor.
$
\end{definition}

\noindent
This value is recorded such that each directed edge $\overrightarrow{wu}$ in $G_b^*$ has access to the value $r_u(w)$. 

\subparagraph{Data Structure.}
Chekuri et al.~\cite{chekuri2024adaptive}  equip the oriented graph $G_b^*$ with a data structure. 
They store for each vertex $u \in V$ the following (see also Figure~\ref{fig:data_types}):
\begin{enumerate}[(a)]
    \item The \emph{exact value} of $d_b^*(u)$. 
    \item The set $N^+(u)$ in a linked list and a pointer some current `position' in the linked list. 
    \item The set $N^-(u)$ in a doubly linked list of buckets $B_j(u)$ sorted by $j$ from high to low.
    Each bucket $B_j(u)$ contains a doubly linked list of all $w \in N^-(u)$ where
    $r_u(w)=j$.
    \item A pointer to the bucket $B_j(u)$ where $j$ is closest to $\log_{1 + \lambda} d^+(u)$. 
\end{enumerate}

\subparagraph{Data Types.}
To guarantee that all data structures have $O(1)$ update time, the proofs in~\cite{chekuri2024adaptive} reference several pointers stored in their data types. We formalise these here:

\begin{adjustwidth}{1cm}{}
\texttt{Bucket} $B_i(v)$ stores a doubly linked list with all \texttt{DiEdges} $\overrightarrow{uv}$ with $i = r_v(u)$. 
\end{adjustwidth}

\begin{adjustwidth}{1cm}{}
\texttt{Vertex} $u$ stores:
\begin{itemize}
    \item a linked list of all \texttt{DiEdges} $\overrightarrow{uv}$ (corresponding to vertices $v \in N^+(u)$). 
     \item a doubly linked list of all non-empty \texttt{Buckets} $B_i(u)$, ordered by $i$.
    \item a pointer to the \texttt{DiEdge} \emph{robin} which represents our `current position' in $N^+(u)$.

    \item a pointer to the \emph{primary bucket} $B_j(u)$ (here $j$ is closest to $\log_{1 + \lambda} d^+(u)$).
        \item the value $d_b^*(u)$.
\end{itemize}

\end{adjustwidth}

\begin{adjustwidth}{1cm}{}
\texttt{DiEdge} $\overrightarrow{uv}$ represents all directed edges $\overrightarrow{uv}$ in $G_b^*$, simultaneously.
It stores:
\begin{itemize}
    \item  two pointers to the \texttt{Vertices} $u$ and $v$, respectively.
    \item  a \emph{counter} counting how many edges $\overrightarrow{uv}$ there are in $G_b^*$,
    \item the value $r_v(u)$,
    \item a pointer to the location of $\overrightarrow{uv}$ in $N^+(u)$. 
    \item a pointer to the bucket $B_{i}(v)$ of $v$ with $i = r_v(u)$, 
    \item a pointer to the location of $\overrightarrow{uv}$ in the linked list in the bucket $B_i(v)$. 
    \item  a pointer to the \texttt{DiEdge} $\overrightarrow{vu}$. 
\end{itemize}

\end{adjustwidth}

\subsubsection{Algorithm Definition}
\label{sssec:algorithmdef}

The algorithm has three parameters that are used throughout its algorithmic subroutines: $\theta \in \{0, 1 \}$, $\lambda \in [0, 1]$ and $b \in \mathbb{N}$. We explain their interplay later in this subsection. 
Each update operation in $G$ simply invokes $b$ updates to the digraph $G_b^*$ (see Algorithm~\ref{alg:insertion_in_G} and \ref{alg:deletion_in_G}). 

\begin{nolinenumbers}
\hspace{-1.1cm}\begin{minipage}[t]{.5 \textwidth}
 \begin{algorithm2e}[H]\internallinenumbers
    \caption{Insert(edge $\{u, v \}$ in $G$)}
    \label{alg:insertion_in_G}
    \begin{algorithmic}
      \FOR{$i \in [b]$ }
      \IF{ $d_b^*(u) \leq d_b^*(v)$}
      \STATE Insert($\overrightarrow{uv}$)
      \ELSE
      \STATE Insert($\overrightarrow{vu}$)
      \ENDIF
      \ENDFOR
    \end{algorithmic}
  \end{algorithm2e}
\end{minipage}~%
\begin{minipage}[t]{.5\textwidth}
\null
 \begin{algorithm2e}[H]
    \nolinenumbers
    \caption{Delete(edge $\{ u, v \}$ in $G$)}
    \label{alg:deletion_in_G}
   \begin{algorithmic}
   \nolinenumbers
      \FOR{$i \in [b]$ }
      \IF{ $u \in N^-(v)$}
      \STATE Delete($\overrightarrow{vu}$)
      \ELSE
      \STATE Delete($\overrightarrow{uv}$)
      \ENDIF
      \ENDFOR
  \end{algorithmic}
  \end{algorithm2e}
\end{minipage}
\end{nolinenumbers}

Recall that the algorithm does not always immediately update the data structure.  
Rather, it only exactly maintains the value $d_b^*(u)$ for each vertex $u$ and the set $N^+(u)$.
The bucketing data structure is updated only when an \emph{information update}  (Algorithm~\ref{alg:information}) is triggered.

 \begin{algorithm2e}[H] \internallinenumbers
 \caption{
 InfoUpdate(\texttt{DiEdge}  $\overrightarrow{xy}$, \texttt{Bucket} $B_j(y)$)}
    \label{alg:information}
    \begin{algorithmic}
      \STATE $r_y(x) = \log_{1 + \lambda} d_b^*(x)$
      \STATE Starting from $B_j(y)$ find the bucket $B_i(y)$ with $i = r_y(x)$ \emph{ \small Inspect at most} $\small O(1)$ \emph{\small buckets}
      \STATE Move $\overrightarrow{xy}$ into the bucket $B_i(y)$
    \end{algorithmic}
  \end{algorithm2e}
  
Suppose that we add a directed edge $\overrightarrow{uv}$ to $G_b^*$. If there already exists an edge $\overrightarrow{uv}$ in $\overrightarrow{G_b}$ then we implicitly insert the extra copy by identifying the corresponding \texttt{DiEdge} $\overrightarrow{uv}$ and incrementing the counter. Otherwise, we need to update $N^+(u)$ and do an information update. The same principle holds when removing a directed edge, where we only do work if we remove the last copy of a directed edge in $G_b^*$ (Algorithm~\ref{alg:add} and~\ref{alg:remove}).

\begin{nolinenumbers}
\hspace{-1cm}\begin{minipage}[t]{.5 \textwidth} 
 \begin{algorithm2e}[H]
\internallinenumbers
    \caption{\mbox{ Add(\texttt{DiEdge} $\overrightarrow{uv}$)}}
    \label{alg:add}
    \begin{algorithmic}
       \STATE Increment the counter of  $\overrightarrow{uv}$.
      \STATE $d_b^*(u) = d_b^*(u) + 1$
      \IF{ 
      $\overrightarrow{uv}$.counter $= 1$
     }
      \STATE Add $\overrightarrow{uv}$ at the end of $N^+(u)$
      \STATE $B_j(v) \gets$ the \emph{primary} \texttt{Bucket} of $v$
      \STATE InfoUpdate($\overrightarrow{uv}$, $B_j(v)$)
      \ENDIF
    \end{algorithmic}
  \end{algorithm2e}
\end{minipage}~%
\begin{minipage}[t]{.5\textwidth}
\null
 \begin{algorithm2e}[H]
    \caption{Remove(\texttt{DiEdge} $\overrightarrow{uv}$)}
    \label{alg:remove}
    \begin{algorithmic}
    \nolinenumbers
    \STATE Decrement the counter of $\overrightarrow{uv}$
   \STATE  $d_b^*(u) = d_b^*(u) - 1$     
      \IF{ 
      $
        \overrightarrow{uv}$.counter $= 0$
     }
      \STATE Remove $v$ from $N^+(u)$
      \STATE $B_i(v) \gets$ the \texttt{Bucket} stored at $\overrightarrow{uv}$
    \STATE InfoUpdate($\overrightarrow{uv}$, $B_i(v)$)
      \ENDIF
    \end{algorithmic}
  \end{algorithm2e}
\end{minipage}
\end{nolinenumbers}

The true complexity of this algorithm lies in the logic surrounding directed edge insertions and deletions. 
Intuitively,  inserting an edge $\overrightarrow{uv}$ iterates over the out-edges $\overrightarrow{ux}$ of $u$ in a round-robin fashion.
If the out-degree of $u$ is considerably larger than the out-degree of $x$, we flip an edge $\overrightarrow{ux}$ to become an edge $\overrightarrow{xu}$, we terminate the round-robin and recurse on the insertion of $\overrightarrow{xu}$.
After this flip, the out-degree of $u$ is unchanged and thus locally, the orientation is still $\lambda$-fair. 
This increased the out-degree of $x$ by one, and we recurse on $x$ (Algorithm~\ref{alg:insertion_in_G}). 
Deleting an edge $\overrightarrow{uv}$ only considers a single directed edge, which is an edge $\overrightarrow{xu}$ where the out-degree $d_b^*(x)$ is approximately maximal. 
If $d^*_b(u)$ is considerably larger than $d_b^*(u)$ then we flip an edge $\overrightarrow{xu}$.
After this flip, the out-degree of $u$ is unchanged. 
Instead, the out-degree of $x$ is decreased by one, and we recurse on $x$ (Algorithm~\ref{alg:insertion_in_G}).

Finally if, after either an edge insertion or deletion, a vertex $u$ actually changes its out-degree then the algorithm will inform up to $\frac{2}{\lambda}$ out-neighbours of $u$ of its new out-degree.

\begin{nolinenumbers}
 \hspace{-1.1cm}   \begin{minipage}[t]{.55 \textwidth}
\null 
 \begin{algorithm2e}[H]
 \internallinenumbers
    \caption{Insert(\texttt{DiEdge} $\overrightarrow{uv}$)  }
    \label{alg:insertion_rank}
    \begin{algorithmic}
    \STATE Add($\overrightarrow{uv}$)
    \FOR{ $\lceil{\frac{2}{\lambda}}\rceil$ iterations do:} 
        \STATE $u$.robin = ($u$.robin + $1$)  mod $|N^+(u)|$
                \STATE \texttt{DiEdge} $\overrightarrow{ux} =$ *$u$.robin 
      \IF{ 
      $
      d_b^*(u) > \max\{(1 + \lambda) \cdot d_b^*(x) + \theta,\frac{b}{4} \}
      $
     }
      \STATE Remove($\overrightarrow{ux})$
      \STATE \emph{Insert($\overrightarrow{xu}$)}
      \STATE \textsc{Break}
      \ENDIF
      \ENDFOR
    \FORALL {
      the next $\lceil{\frac{2}{\lambda}}\rceil$ $\overrightarrow{uw}$ in $N^+(u)$ 
    }
    \STATE InfoUpdate($\overrightarrow{uw}$, $\overrightarrow{uw}$.Bucket)
    \ENDFOR

    \end{algorithmic}
  \end{algorithm2e}
\end{minipage}~%
\begin{minipage}[t]{.5\textwidth}
\null
 \begin{algorithm2e}[H]
    \caption{Delete(\texttt{DiEdge} $\overrightarrow{uv}$)}
    \label{alg:deletion_rank}
    \begin{algorithmic}
    \nolinenumbers
     \STATE Remove($\overrightarrow{uv}$)
      \STATE $x \gets \textnormal{First( Max( Bucket($N^-(u)$)))} $
      \IF{
      $
       d_b^*(x) > 
      \max \{(1 + \lambda) \cdot d_b^*(u)   +  \theta,{\frac{b}{4}} \}
      $
}
            \STATE Add$(\overrightarrow{ux})$
            \STATE \emph{Delete($\overrightarrow{xu}$})
      \ELSE
  \FOR {the next $\lceil{\frac{2}{\lambda}}\rceil$ $\overrightarrow{uw}$ in $N^+(u)$ }
      \STATE InfoUpdate($\overrightarrow{uw}$, $\overrightarrow{uw}$.Bucket)
   \ENDFOR
     \ENDIF
  \end{algorithmic}
  \end{algorithm2e}
\end{minipage}
\end{nolinenumbers}

\subparagraph{Algorithm Parameters and Asymptotic Complexity.}
The algorithm has three parameters:
\begin{itemize}
    \item $\theta \in \{ 0, 1 \}$ is a Boolean that is used in the logic of Algorithms~\ref{alg:insertion_rank} and \ref{alg:deletion_rank}. The Boolean specifies whether we want a slower $(1+\eps)$-approximation (setting $\theta \gets 0$), or, whether we want a faster algorithm that achieves an additive $O(\log n)$-error (setting $\theta \gets 1$). 
    \item $\lambda \in [0, 1]$ is the approximation parameter. 
    \begin{itemize}
        \item If $\theta = 0$, then we may choose $\lambda \in \Theta(\frac{\eps^2}{\log n})$ to obtain a $(1+\eps)$-approximation~\cite{chekuri2024adaptive, christiansen2025local}. 
        \item If $\theta = 1$, then we may choose $\lambda \in \Theta(\frac{1}{\log n})$ to obtain an additive $O(\log n)$ error~\cite{chekuri2024adaptive, christiansen2025local}. 
    \end{itemize}
    \item $b \in \mathbb{N}$ records how many edges the multigraph $G_b$ has per edge in $G$. If $\theta = 1$ then $b = 1$.
    
    If $\theta = 0$ then $b > \frac{4}{\lambda}$ in order to make the algorithmic logic work. 
    Indeed, consider an edge $\{ u, v \}$ in $G$. 
    Our algorithm maintains the invariant that if $\overrightarrow{uv}$ is in $G_b^*$ then $d^*_b(u) \leq (1+\lambda)d^*_b$. If both $d^*_b(u)$ and $d^*_b(v)$ are less than $\frac{1}{\lambda}$ then there is no way to orient $\{ u, v \}$ in $G_b^*$ such that our invariant is satisfied. 
\end{itemize}

Each update in $G$ triggers at most $b$ updates in $G_b$.
Each update to $G_b$ calls Algorithm~\ref{alg:insertion_rank} or \ref{alg:deletion_rank}. 
These functions do at most $O(\frac{1}{\lambda})$ constant-time operations before recursing.
These functions recurse at most $O(\log_{1+\lambda} n)$ times. 
It follows that the total runtime is at most:
\[
O( b \cdot \frac{1}{\lambda} \cdot \log_{1 + \lambda} n) \subset O(b \lambda^{-2} \log n) \Rightarrow   \begin{cases}
     O( \lambda^{-3} \log n)  & \textnormal{ if } \theta = 0 \\
     O( \lambda^{-2} \log n) &  \textnormal{ if } \theta = 1
   \end{cases} 
\]

In theory, one chooses $\lambda \in \Theta(\eps^2 / \log n)$ where $\eps$ is the approximation constant.
In practice, we simply choose a smaller $\lambda$ for a better approximation.

\subparagraph{Space Usage.}
In theory, the data type as we specified in this paper allows each of the operations in Algorithms~\ref{alg:insertion_in_G}--\ref{alg:deletion_rank} to take constant time. 
We give three examples:  (i) we can flip a directed edge in $G_b^*$ in constant time  because the \texttt{DiEdge} $\overrightarrow{ux}$ stores a pointer to the  \texttt{DiEdge} $\overrightarrow{xu}$, (ii) a \texttt{DiEdge} $\overrightarrow{uv}$ can remove itself from the linked list $N^+(u)$ in constant time because it stores a pointer to its location, and (iii) a  \texttt{DiEdge} $\overrightarrow{uv}$ can remove itself from the doubly linked list $B_i(v)$ of $N^-(v)$ in constant time because it stores a pointer to its location.

While theoretically efficient, this solution faces a practical bottleneck due to high space usage. Indeed, it stores six pointers and two integers per edge in the original graph $G$. 

Moreover, the excessive use of pointers makes executing the algorithm less predictable and slowing it down in practice. Additionally, each pointer is not efficient in addressing only a part of the memory, i.e. the out-going edges, since it requires constant (64-bit) memory. 

\section{Engineering Aspects}
\label{sec:improvement}

We now engineer an efficient implementation of~\cite{chekuri2024adaptive}.
Our key observation is that there are two data containers where the order of the elements in the container is irrelevant:
\begin{itemize}
    \item \texttt{Vertex} $u$ stores all $\texttt{DiEdges}$ $\overrightarrow{uv}$ in arbitrary order, and
    \item \texttt{Bucket} $B_i(v)$ stores all $\texttt{DiEdges}$ $\overrightarrow{uv}$ with $i = r_v(u)$ in arbitrary order.
\end{itemize}
By sorting these sets as arrays, we obtain more control over the size of these objects and the size of pointers to elements in these sets.
This in turn allows us to use packing techniques to store the same information in fewer bits.

\subparagraph{Arrays.}
This allows us to use arrays instead of linked list as we give each: 

\begin{itemize}
    \item \texttt{Vertex} $u$
an array $A^+_u$ with all \texttt{DiEdges} $\overrightarrow{uv}$ in arbitrary order.
\item \texttt{Bucket} $B_i(v)$ an array $B_i^v$ with all \texttt{DiEdges} $\overrightarrow{uv}$ with $i = r_v(u)$ in arbitrary order.
\end{itemize}

Adding a \texttt{DiEdge} $\overrightarrow{uv}$, given the \texttt{Vertex} $u$ and the correct \texttt{Bucket} $B_i(v)$, takes amortized constant time. Indeed, we simply increase the sizes of $A^+_u$ and $B_i^v$ and append $\overrightarrow{uv}$. 
Removing $\overrightarrow{uv}$ also takes amortized constant time: given its index $a$ in $A^+_u$ (and $b$ in $B_i^v$) we may swap $A^+_u[a]$ with the last element of $A^+_u$ (swap $B_i^v[b]$ with the last element) and decrement the array's size by one. 
These operations are supported by the \texttt{vector} data type in C++.

\subparagraph{From Arrays to Packing.}
Replacing doubly linked lists with vectors naturally reduces algorithmic overhead. It also has a secondary benefit: instead of maintaining, for a \texttt{DiEdge} $\overrightarrow{uv}$, pointers to its location in linked lists we store integers maintaining its index in vectors. Recall that we maintain an orientation $G_b^*$ where the maximum out-degree $\texttt{Max} (G_b^*)$ is low. 
As a consequence, for any vertex $v$, the array $A^+_u$ is small. Thus, for a \texttt{DiEdge} $\overrightarrow{uv}$ its index in $A^+_u$ is small. 
Via the same logic, the integers \emph{robin}, $d^*_u(u)$, $\lfloor \log_{1+\lambda} (r_u(v)) \rfloor$ and the index of the \texttt{DiEdge} $\overrightarrow{vu}$ are small. 
We pack these small values into 32-bit integers. 

\subparagraph{Indirection.}
Any \texttt{DiEdge} $\overrightarrow{uv}$ contained pointers to: the \texttt{Vertex} $u$, the \texttt{Vertex} $v$, and the \texttt{DiEdge} $\overrightarrow{vu}$. 
If each $\texttt{DiEdge}$ only stores a pointer to its target then we can access the \texttt{Vertex} $u$ from $\overrightarrow{uv}$ by  first accessing the \texttt{DiEdge} $\overrightarrow{vu}$ and then reading its target.

\subparagraph{Removing Bucket Pointers.}
Each \texttt{Vertex} $u$ maintained a pointer to the primary bucket $B_j(u)$ and each \texttt{DiEdge} $\overrightarrow{uv}$ maintained a pointer to the bucket $B_i(v)$ that contains $\overrightarrow{uv}$. 
These pointers are only used in  Algorithm~\ref{alg:information}. Chekuri et al.~\cite{chekuri2024adaptive} prove that to find for a \texttt{DiEdge} $\overrightarrow{xy}$ is current bucket, one steps over at most $O(1)$ buckets from the given pointer. We choose to remove these pointers. Instead, we find the correct bucket for $\overrightarrow{xy}$ by looping over all buckets $B_i(y)$. 
Observe that there are at most $O(\log_{1+\lambda} n) \subseteq O(\lambda^{-1} \log n)$ buckets $B_i(y)$. Moreover, a call to Algorithm~\ref{alg:insertion_rank} or~\ref{alg:deletion_rank} triggers $O(\lambda^{-1})$ information updates \emph{only if} it does not recurse. 
So, the asymptotic running time remains the same.

\begin{figure}[b]
    \centering
    \includegraphics[page = 2]{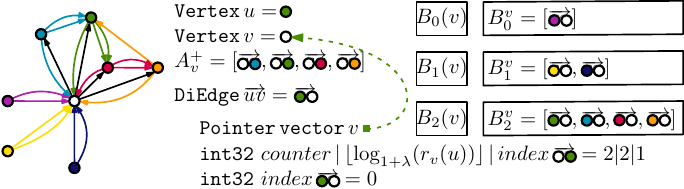}
    \caption{We illustrate our more compact data structure compared to the one from Figure~\ref{fig:data_types}. 
    }
    \label{fig:compact}
\end{figure}

\subparagraph{Compact Types.}
By our above analysis, we obtain more compact data types (see Figure~\ref{fig:compact}):

\begin{adjustwidth}{1cm}{}
\texttt{Bucket} $B_i(u)$ stores a vector $B_i^u$ with \texttt{DiEdges} $\overrightarrow{wu}$ with $i = r_u(w)$. 
\end{adjustwidth}

\begin{adjustwidth}{1cm}{}
\texttt{Vertex} $u$ stores:
\begin{itemize}
    \item a vector $A_u^+$ storing all \texttt{DiEdges} $\overrightarrow{uv}$ (corresponding to vertices $v \in N^+(u)$). 
     \item a  linked list of all non-empty \texttt{Buckets} $B_i(u)$, ordered by $i$.
    \item a 32-bit integer storing our `current position' \emph{robin} in $A^+_u$  and the value $d_b^*(u)$.
\end{itemize}
\end{adjustwidth}

\begin{adjustwidth}{1cm}{}
\texttt{DiEdge} $\overrightarrow{uv}$ represents all directed edges $\overrightarrow{uv}$ in $G_b^*$, simultaneously.
It stores:
\begin{itemize}
    \item  A pointer to the \texttt{Vertex} $v$.
    \item A 32-bit integer storing the \emph{counter}, the value $r_v(u)$, and the index of $\overrightarrow{vu}$ in $A^+_u$. 
    \item An 32-bit integer storing the index of $\overrightarrow{uv}$ in $B_i^v$ for $i = r_v(u)$. 
    \end{itemize}
    
\end{adjustwidth}

\subsection{A Space-Time Trade-off}

Finally, we propose one more modification to the algorithm.
The algorithm in~\cite{chekuri2024adaptive} and our above implementation store for a vertex $v$ all non-empty buckets $B_i(v)$ in a  linked list.
The algorithm in~\cite{chekuri2024adaptive} accesses these buckets by storing a pointer for each edge in $G$. 
In our implementation, we find the bucket $B_i(v)$ where $i = r_v(u)$ in $O(\log_{1+\lambda} n)$ time by iterating over all buckets $B_j(v)$. We showed that that this does not increase the asymptotic running time.
However, it does increase the \emph{actual} running time of the program.

Both methods have an inherent downside:~\cite{chekuri2024adaptive} uses a lot of memory per edge, and our above implementation is relatively slow. 
Moreover, doubly linked list inherently bring a lot of overhead and cache misses. 
We therefore propose a third solution to store the buckets:

We define for a \texttt{Vertex} $v$ the \emph{Bucket array} $\mathbb{B}_v$ such that for all $i$, $\mathbb{B}_v[i] = B_i(v)$. 
Any \texttt{DiEdge} $\overrightarrow{uv}$ can access the bucket $B_i(v)$ where $i = r_v(u)$ by simply indexing $\mathbb{B}_v[r_v(u)]$ in constant time. 
By storing the array $\mathbb{B}_v$ as a vector, we require only a relatively small amount of memory for every empty bucket.
There are $O(\log_{1+\lambda}(n) )$ buckets for each vertex $v$, so this additively increases our space usage by $O( \frac{n}{\lambda} \log n)$.
Since the number of edges $m$ is considerably larger than the number of vertices $n$, this is space-efficient when compared to~\cite{chekuri2024adaptive}. 
This solution does use more space compared to our above implementation, with the benefit of having a significantly faster implementation of Algorithm~\ref{alg:information}. 
Thus, we offer two implementations -- one using singly linked lists, and one using this array data structure -- that together provide a trade-off between time and space used.

\section{Experimental Evaluation}
\label{sec:experiments}
In this section we evaluate our algorithm and compare its performance with previous work by Großmann~et~al.~\cite{grossmann2025engineering} and Borowitz~et~al.~\cite{dblp:conf/acda/borowitzg023}.
We compare them on the sparse data set introduced in \cite{dblp:conf/acda/borowitzg023} and present a new dense data set featuring also deletions.
The analysis is done in separate sections (sparse, dense) since the performance of the various algorithms highly depend on the density of used inputs.
\subparagraph{Methodology.}
We carried out the implementation of the algorithms in \cpp and compiled the programs with \textsf{g}\texttt{++-}\textsf{11.4}.
We tested each algorithm on a machine equipped with 768 GB of RAM and an AMD EPYC 9754 128-Core Processor running at 2.25 GHz having a cache of 256 MB.
The time loading the graphs is not measured, but instead we measure every operation and sum up the timings.
We repeated every experiment 3 times, computations taking longer than a day in total were cancelled and are marked as out of time (OOT) in the result tables.
For comparing the results we use performance profiles as introduced by Dolan and Moré~\cite{dblp:journals/mp/dolanm02}. For a factor of $\tau>1$ we plot the fraction of instances that could be solved within that factor of the best result per instance. Algorithms in the top left corner of the plot (e.g. Figure~\ref{fig:quality}) are considered the best algorithm. 
\subparagraph{Instances.}
We use two sets of graphs for the experiments.
First, the \emph{sparse set} containing 83 instances collected by Borowitz~et~al.~\cite{dblp:conf/acda/borowitzg023}.
Only four of the instances feature deletions.
The size of the graphs is small to medium and they have low-density and out-degree.
Second, the \emph{dense set}, that contains three, large, denser graphs sourced from the Suite Sparse Matrix collection~\cite{Davis2011UFSparse} and four ultra-dense graphs generated with KaGen~\cite{funke2017communication}. See our supplement details.
The ultra-dense graphs were generated in the Erdos-Renyi model having 8192 vertices and between $\frac{1}{2}$ and $\frac{7}{8}$ of all possible edges.  
The graphs and their features are listed in Table~\ref{tab:graph_properties} in Appendix~\ref{app:features:graph}.
For these instance we generated three different update sequences. The first, building up the graph in lexicographical ordering (\textsc{Lex}). The other two also include deletion of edges afterwards.
The edges to be deleted are selected based on an optimal orientation of the final graph, computed with HeiOrient~\cite{reinstadtler2024heiorient}.
In the \textsc{Lex+Top50} configuration all edges outgoing from the upper half of vertices ordered by out-degree are deleted.

The \textsc{Lex+Out50} configuration is built by maintaining a heap of the vertices by out-degree.
Outgoing edges are deleted and the out-degree is updated in the heap until the maximum out-degree is half of the optimal orientation.
We select these sequences in order to see how different deletion strategies impact the quality, memory and runtime metrics.

\subparagraph{Comparisons and Implementations.}
We refer to our basic implementation of the algorithm  by Chekuri et al.~\cite{chekuri2024adaptive} as \algorithmCCHHQRS and 
with our engineering improvements 
as \algorithmPCCHHQRS. 
We provide two implementations of \algorithmPCCHHQRS.
One with \algorithmPCCHHQRS \textsf{list} organizes the in-buckets in a singly linked list, and the one without suffix works with directly addressable, possibly empty, buckets.
We compare with \competitorLimitedBFS by Borowitz~et~al.~\cite{dblp:conf/acda/borowitzg023} and \competitorStrongBFS and  \competitorImprovedBFS by Gro{\ss}mann~et~al.~\cite{grossmann2025engineering}.
These algorithms are the best performing in the extensive studies in \cite{dblp:conf/acda/borowitzg023} and \cite{grossmann2025engineering} with regards to the max out-degree metric $\max(\overrightarrow{G})$.
The sum$^2(\overrightarrow{G})$ metric has not been studied in practice.
The \competitorLimitedBFS algorithm performs a limited breadth-first-search of length up to 20 to find improving paths upon insertion and removes only the edge on deletion, not optimizing the orientation.
The \competitorStrongBFS runs a breadth-first-search to find improving paths on every insertion and deletion, returning the optimal orientation concerning out-degree and squared-sum of out-degree. The \competitorImprovedBFS only searches for improving paths if it is necessary to maintain the optimal out-degree.
\subparagraph{Parameters.}
The algorithms \algorithmCCHHQRS and \algorithmPCCHHQRS derived in Section~\ref{sub:chekuri} and Section~\ref{sec:improvement} have the three parameters: $\lambda$, $\theta$, $b$.
First, $\lambda>0$ and $\theta\in\{0,1\}$ control the quality of the approximation by defining its $\lambda$-fairness.
If $\theta=0$, $b$ must be larger than~$\frac{1}{\lambda}$ so that the fractional orientation inequality can be satisfied.
Otherwise ($\theta=1$), $b$ can be set to 1 making the orientation integral. This comes at costs of quality in the form of an additive error of $O(\log n)$ in the approximation.
\subsection{Experiments on Sparse Instances}
The following experiments are conducted on the sparse instance set collected by Borowitz~et~al.~\cite{dblp:conf/acda/borowitzg023} and analyse the running times and quality of the orientations. 
Since some of graphs in this set were too small to consistently measure the maximum resident memory by repeatedly querying the linux internal process stats, we examine the memory consumption only for the dense instances.
The maximum peak memory is 23 GB for the \algorithmCCHHQRS and \algorithmPCCHHQRS with $\lambda=0.01$ and $\theta=0$ implementations on the \textsc{ljournal-2008} instance.
The \algorithmPCCHHQRS~\textsf{list} algorithm with $\lambda=0.1$, $\theta=1$ only requires 4.4 GB on this instance - only 2.29 times the amount of \competitorStrongBFS.
\begin{figure}
\centering
\includegraphics[width=\textwidth]{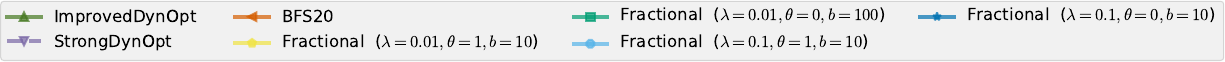}

\begin{subfigure}[t]{0.49\textwidth}
\centering
\includegraphics[width=\textwidth]{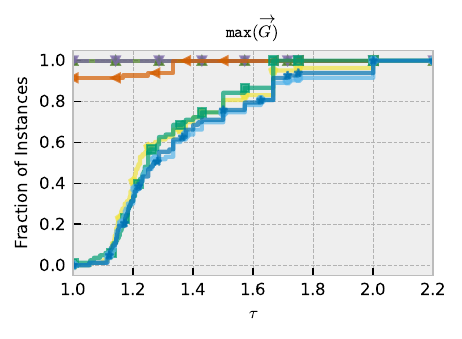}
\label{subfig:outdegree:prevwork}
\end{subfigure}
\begin{subfigure}[t]{0.49\textwidth}
\centering
\includegraphics[width=\textwidth]{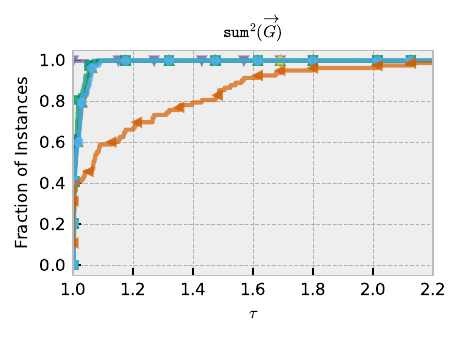}
\label{subfig:quality:prevwork}
\end{subfigure}
\caption{State-of-the-art comparison with different quality measures on the sparse instance set. The \textsf{PackedF*} is omitted due to its similar results as \algorithmCCHHQRS with a maximum difference of~0.3\%.} 
\label{fig:quality}

\end{figure}

\subparagraph{Quality Metrics.} In Figure~\ref{fig:quality} we compare different configurations of \algorithmCCHHQRS with the state-of-the-art for integral orientations. We evaluate the performance using the two quality measures discussed in Section~\ref{sec:prelim}. 
For better readability, the results for the packed implementations are omitted. They achieve similar to those results of the \algorithmCCHHQRS configurations with a maximum difference in quality of less than 0.3\%.
We can observe that smaller values of $\lambda$ lead to improved solution quality for both metrics.

\textit{Maximum Out-Degree.} When comparing the algorithms using the metric $\max(\overrightarrow{G})$, we can observe that
both exact approaches \competitorImprovedBFS and \competitorStrongBFS always find the optimum solution. 
Compared to these, \algorithmCCHHQRS performs worse overall, but maintains a factor of less than two for all parameter choices.

\textit{Squared Sum of Out-Degrees.} When evaluating with sum$^2(\overrightarrow{G})$ of the naively rounded fractional orientations, \algorithmCCHHQRS performs significantly better than the \competitorLimitedBFS algorithm, matching the results of \competitorImprovedBFS.
The \competitorStrongBFS algorithm returns the best values on every instance.
This is the expected behaviour, because improving all possible paths minimizes both objective functions.
Our newly engineered algorithms perform very well in the $\mathrm{sum}^2(\overrightarrow{G})$ metric as they were designed for it.
The variation in $\lambda\in\{0.01,0.1\}$ is not visible in this metric.
In exchange, they allow themselves a higher maximum out-degree on some of the vertices which is not accounted for in the definition of $\lambda$-fair.
\begin{figure}
    \centering
        \input{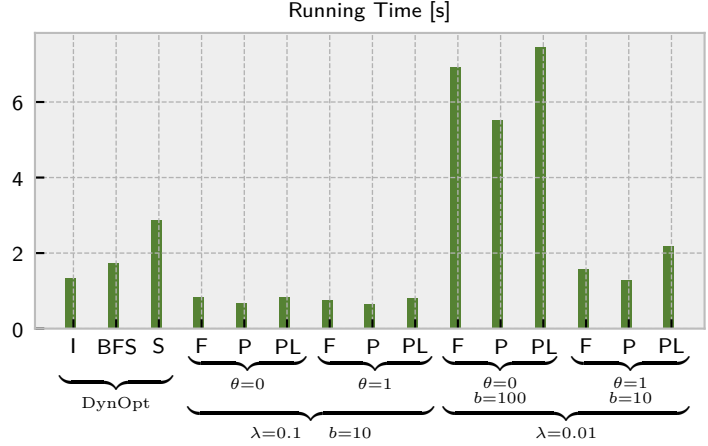}
    \caption{Running times on the sparse set comparing \algorithmCCHHQRS (F), \algorithmPCCHHQRS (P), and \algorithmPCCHHQRSList (PL) with \competitorLimitedBFS (BFS), \competitorImprovedBFS (I), and \competitorStrongBFS (S).
    Parameters: $\lambda$ overall fairness of the orientation maintained. $\theta$ is a binary variable deciding an additional $O(\log{n})$ error, $b$ is the number of fractional edges.}
    \label{fig:results:runtime:sparse:plot}
\end{figure}

\subparagraph{Running Time.} Figure~\ref{fig:results:runtime:sparse:plot} shows the geometric mean running time for the approaches  on the sparse data set. The normalized running times can be found in Tables~\ref{tab:running:time:prevwork1} and \ref{tab:running:time:prevwork2} in the Appendix.
Our fastest algorithm \algorithmPCCHHQRS is twice as fast as \competitorImprovedBFS, which is the fastest state-of-the-art method on this data set. The running time of our approach is highly dependent on the parameter choices.
Setting $\theta=1$ is faster than $\theta=0$. Decreasing $\lambda$ results in longer running times. The smallest value $\lambda=0.01$ we tested is 8.58 times slower than the fastest configuration with $\lambda=0.1$.
The packed list implementation is up to  35\,\% slower.
The measured difference between \competitorImprovedBFS and \competitorLimitedBFS matches the result in \cite{grossmann2025engineering}.


\subsection{Experiments on Dense Instances}
We conduct the following experiments on the new dense instance set, which includes three real-world graphs and four ultra-dense graphs. As in previous sections, we analyze running time and solution quality, and additionally provide an in-depth examination of \hbox{memory usage.}
\subparagraph{Quality Metrics.}
Only three out of the 21 instances were solved optimally by \competitorStrongBFS on the set of dense instances in the given timeout of 24 hours. Therefore, we show  performance profiles for our parameter choices and \algorithmPCCHHQRS in tour parameter choices and \algorithmPCCHHQRS in Figure~\ref{fig:res:21} on all instances and  limited the following  discussion about quality on those three~instances.

\textit{Maximum Out-Degree.} Evaluating the performance using $\max(\overrightarrow{G})$ the solutions of \algorithmCCHHQRS are between \numprint{24}\,\% ($\lambda=0.01$) up to \numprint{30}\,\% ($\lambda=0.1$) worse than the optimal orientations.
The method \competitorLimitedBFS also performs worse on these instances with solutions being \numprint{26}\,\% worse. Note that this approach does not handle deletions explicitly by running a improving path search, which can explain the observed suboptimal results on our instances featuring deletions. 
These results match the observations of previous work.

\textit{Squared Sum of Out-Degrees.} Using sum$^2(\overrightarrow{G})$, we observe a solution quality for \algorithmPCCHHQRS which is less than \num{0.8}\,\% higher for $\lambda=0.1$ and only \num{0.3}\,\% higher when setting $\lambda=0.01$ than the optimum computed by \competitorStrongBFS. With this, we achieve better results on these instances compared to  \competitorImprovedBFS, which in the geometric mean has a solution quality that is \numprint{15}\,\% worse.
The \competitorImprovedBFS is specialised in maintaining the minimum maximum out-degree while requiring fewer operations than \competitorStrongBFS. 
Therefore, its sum$^2(\overrightarrow{G})$ metric is worse than our results. 
The impact of the $\lambda$ factor, in the range we tested, on the sum$^2(\overrightarrow{G})$ metric  is small for our new algorithms.
 \begin{figure}
    \centering        \input{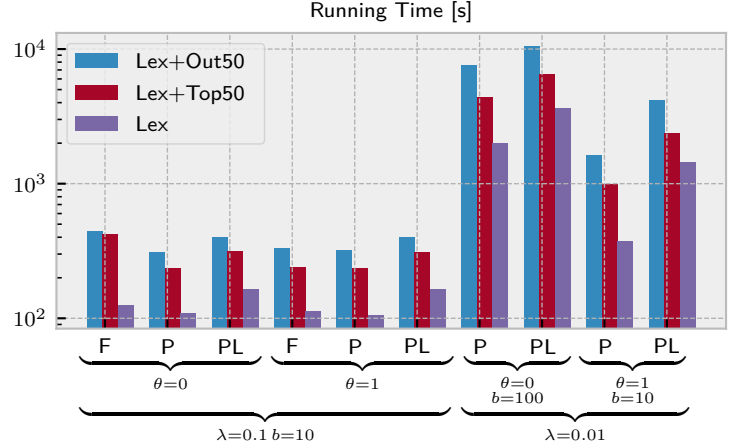}
    \caption{Geometric mean running times on the dense set comparing \algorithmCCHHQRS (F), \algorithmPCCHHQRS (P), and \algorithmPCCHHQRSList (PL). We evaluate them  for the different  deletion sequences.}
    \label{fig:results:runtime:dense}
\end{figure}

\subparagraph{Running Time.}
The running times on the dense set for our implementations are displayed in Figure~\ref{fig:results:runtime:dense} and  Tables~\ref{tab:running:time:prevwork1} and ~\ref{tab:running:time:prevwork2}  of the Appendix.

We detail these results in the full version.
Detailed results per instance are shown in Table~\ref{tab:results:runtime}.
The choice of lambda greatly impacts the running time.
 Reducing the value of $\lambda$ by a factor of 10 from $0.1$ to $0.01$ (keeping $\theta=0$) can result in running time that is  20 times slower in the geometric mean.
Setting $\theta=1$ speeds up the algorithm by up to a factor of 5 for small $\lambda=0.01$ compared to $\theta=0$.
The methods from Borowitz~et~al.~\cite{dblp:conf/acda/borowitzg023} and Großmann~et~al.~\cite{grossmann2025engineering} find solutions for considerably fewer number of instances in this data set while requiring more running time. More precisely, \competitorLimitedBFS, \competitorImprovedBFS, and \competitorStrongBFS find solutions on 15, 5, and 3 instances while having a geometric mean running time which is 112.17, 29.15, and 25.98 times slower than our fastest algorithm (\algorithmPCCHHQRS) on the respective instances.

\begin{figure}
    \centering
    \input{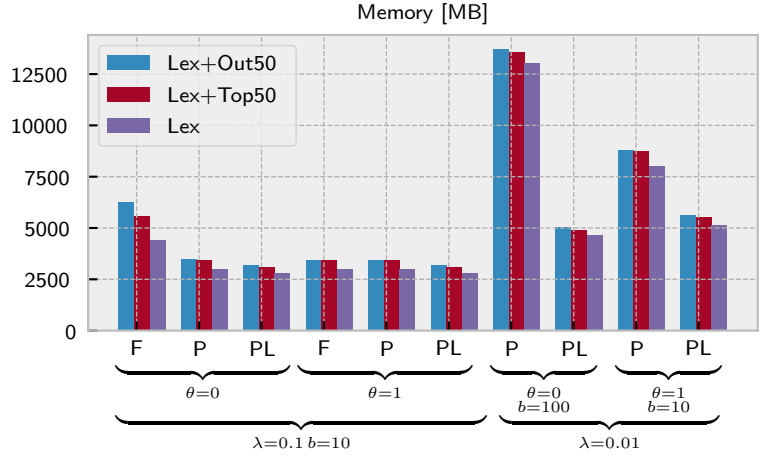}
    \caption{Memory consumption of \algorithmCCHHQRS (F), \algorithmPCCHHQRS (P), and \algorithmPCCHHQRSList (PL) for different configurations evaluated separately for the different deletions sequences.}
    \label{fig:mem_catbar}
\end{figure}

\subparagraph{Memory.}
 As a smaller $\lambda$ value results in more buckets, we expect the memory consumption to increase with decreasing $\lambda$. This behaviour can be observed in Figure~\ref{fig:mem_catbar}.
 Here, we compare the memory consumption of our implementations, which is evaluated separately for the different deletion sequences. 
 Already for $\lambda=0.1$, we can observe that the \algorithmPCCHHQRS is improving over \algorithmCCHHQRS, where we achieve a geometric mean reduction in memory consumption by a factor of 1.94.
 In the case of $\lambda=0.01$, we omit \algorithmCCHHQRS due to its already high memory consumption for $\lambda=0.1$ and focus on comparing the implementations \algorithmPCCHHQRS and \algorithmPCCHHQRSList.
 With this configuration, the \textsf{List} implementation can save up to a factor of 2.72 memory ($\lambda=0.01, \theta=0, b=100$). Overall, we can see that the choice for $\lambda$ has the highest impact regarding memory consumption. However, the parameters $\theta$ and $b$ also influence the memory needed.
 The memory consumption for the competitor algorithms is generally smaller. Compared to our best approach, the competitor methods perform up to a factor of 3.32 better. However, this result is expected since they do not maintain a fractional orientation but an integral one.
 They store each edge's target on both sides in combination with the position of the reverse edge in the other array.
Lastly, we observe that the deletion of edges (\textsc{Top50}, \textsc{Out50}) requires more memory than just the \textsc{Lex} ordering, with the \textsc{Out50} requiring the most.
This is caused by the spreading of indices in the in-bucket, as seen by the smaller difference in \hbox{the \textsf{list} implementations.}

\begin{tcolorbox}[width=\linewidth, colback=white!95!black,left=4pt,right=4pt,top=4pt,bottom=4pt,enlarge top by=-1pt,%
  enlarge bottom by=-1pt]
  \textbf{Practitioner Advice:}
  When optimizing the sum of squared out-degrees we recommend using the packed \algorithmPCCHHQRS implementation since it is the fastest implementation overall and the difference in quality to the exact method (\competitorStrongBFS) is minimal.
  The list implementation should only be used if memory consumption is of utmost importance.
  The choice of parameters is paramount.
  Choosing higher values for $\lambda$ speeds up the computation and reduces the memory needed. Setting $\theta=1$ can be used to relax the running time further, but is penalized by a quality trade-off. 
  The $b$ parameter should be chosen small to speed up computation, but has to stay above $\frac{1}{\lambda}$ if $\theta=0$.
 \end{tcolorbox}

\section{Conclusion}
We have bridged the gap between theoretical results and practical experiments for adaptive edge orientations.
We engineered a space efficient implementation  of a $\lambda$-fair fractional orientation algorithm by Chekuri~et~al.~\cite{chekuri2024adaptive} and presented several effective methods to reduce the memory footprint of the algorithm.
In extensive experiments we show the feasibility of our approach in comparison with several competitors from previous work.
Our results for the squared sum degree metric are close to optimal, while the running time is considerably lower than competing algorithms.
Our algorithm \algorithmPCCHHQRS solves more instances than the competitors within one day of compute time per instance while it is up to 112 times faster on the instances that are solvable by all methods compared.
Our list-based bucket implementations can decrease the memory consumption by a factor of \num{2.72}.
Future avenues of work include reducing the memory costs further and looking at batch updates.
 These aspects could be combined with efforts to parallelize the algorithms.
\bibliographystyle{plainurl}
\bibliography{refs}
\appendix
\newpage
m

\section{Additional Results}
\begin{table}[h]

\caption{Geometric mean and normalized running times (in seconds) on the sparse data set.}
\label{tab:running:time:prevwork1}
\centering
\begin{tabular}{llrrrrr}
\toprule
\bfseries Name & $\lambda$&$\theta$&$b$& \bf $t_{geomean}$ & \bf  $t_{norm}$\\\midrule
    \algorithmPCCHHQRS  &0.1&1&10 & 0.64 & 1.00\\
\algorithmPCCHHQRS  &0.1&0&10 & 0.68 & 1.06\\
\algorithmCCHHQRS  &0.1&1&10 & 0.77 & 1.20\\
\algorithmPCCHHQRSList  &0.1&1&10 & 0.80 & 1.24\\
\algorithmCCHHQRS  &0.1&0&10 & 0.82 & 1.28\\
\algorithmPCCHHQRSList  &0.1&0&10 & 0.84 & 1.31\\
\algorithmPCCHHQRS  &0.01&1&10 & 1.27 & 1.98\\
\competitorImprovedBFS &&& & 1.35 & 2.09\\
\algorithmCCHHQRS  &0.01&1&10 & 1.57 & 2.44\\
\competitorLimitedBFS &&& & 1.73 & 2.69\\
\algorithmPCCHHQRSList &0.01&1&10 & 2.20 & 3.42\\
\competitorStrongBFS &&& & 2.88 & 4.48\\
\algorithmPCCHHQRS  &0.01&0&100 & 5.52 & 8.58\\
\algorithmCCHHQRS  &0.01&0&100 & 6.92 & 10.76\\
\algorithmPCCHHQRSList  &0.01&0&100 & 7.46 & 11.60\\
\bottomrule
\end{tabular}
\end{table}
\begin{table}[h]
\caption{Geometric mean and normalized running time (in seconds) on the 21 dense instances. The competitor methods are omitted since they only find solutions on 15 (\competitorLimitedBFS), 5 (\competitorImprovedBFS), and 3 (\competitorStrongBFS) instances.}
\label{tab:running:time:prevwork2}
\centering
\begin{tabular}{lrrrrr}
\toprule
\bfseries Name & $\lambda$&$\theta$&$b$& \bf $t_{geomean}$ & \bf  $t_{norm}$ \\\midrule
\algorithmPCCHHQRS  &0.1&1&10 & 199.35 & 1.00\\
\algorithmPCCHHQRS  &0.1&0&10 & 199.71 & 1.00\\
\algorithmCCHHQRS  &0.1&1&10 & 208.31 & 1.04\\
\algorithmPCCHHQRSList  &0.1&1&10 & 273.55 & 1.37\\
\algorithmPCCHHQRSList  &0.1&0&10 & 274.74 & 1.38\\
\algorithmCCHHQRS  &0.1&0&10 & 286.34 & 1.44\\
\algorithmPCCHHQRS  &0.01&1&10 & 843.64 & 4.23\\
\algorithmPCCHHQRSList  &0.01&1&10 & 2434.94 & 12.21\\
\algorithmPCCHHQRS  &0.01&0&100 & 4065.26 & 20.39\\
\algorithmPCCHHQRSList  &0.01&0&100 & 6289.29 & 31.55\\
\bottomrule

\end{tabular}
\end{table}
\begin{figure}
    \centering
    \includegraphics{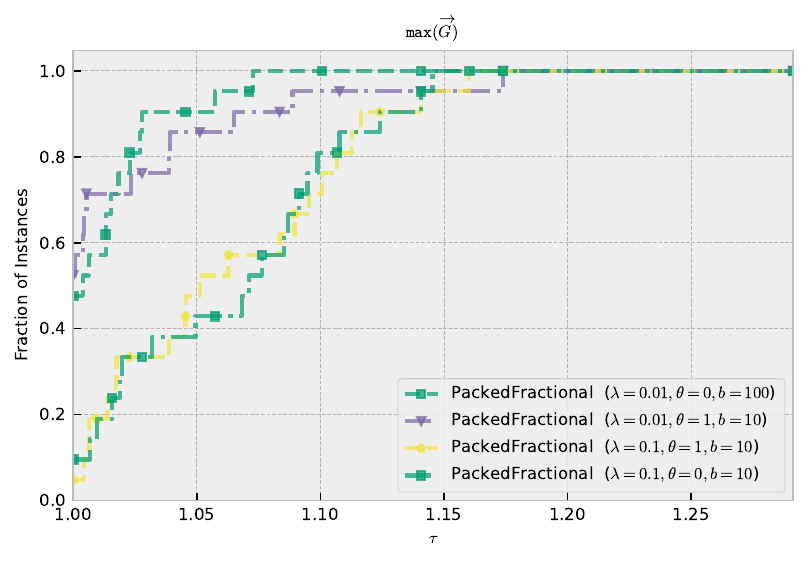}
    \includegraphics{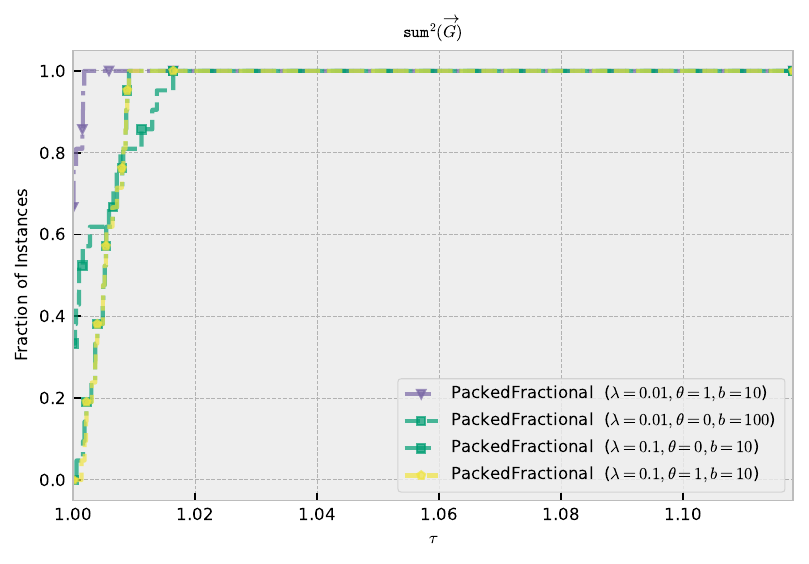}
    \caption{Performance profiles of our \algorithmPCCHHQRS for different choices of $\theta$, $\lambda$ and $b$  on the 21 dense instances. Due to its similar results we omit \algorithmCCHHQRS. The competitors by Großmann~et~al.\cite{grossmann2025engineering} and Borowitz~et~al.~\cite{dblp:conf/acda/borowitzg023} did not solve all instances in time and are omitted.}
    \label{fig:res:21}
\end{figure}
\setlength{\tabcolsep}{3pt}
\begin{sidewaystable}[]
    \centering
    {
    \begin{tabular}{llrrrrrrrrrrrrr}
    \toprule
    &&\multicolumn{3}{c}{Competitors}&\multicolumn{6}{c}{$\lambda=0.1$ $b=10$}&\multicolumn{4}{c}{$\lambda=0.01$}\\\cmidrule(lr){3-5}\cmidrule(lr){6-11}\cmidrule(lr){12-15}
    &&\multicolumn{3}{c}{}&\multicolumn{3}{c}{$\theta=0$}&\multicolumn{3}{c}{$\theta=1$}&\multicolumn{2}{c}{$\theta=0$,$b=100$}&\multicolumn{2}{c}{$\theta=1$,$b=10$}\\\cmidrule(lr){6-8}\cmidrule(lr){9-11}\cmidrule(lr){12-13}\cmidrule(lr){14-15}
Instance&Mode&BFS20&Im.&Strong&F&P&PL&F&P&PL&P&PL&P&PL\\\midrule
gnm-undir'-16777216&\textsc{Out50}&59769.82&\em OOT&\em OOT&205.69&\bfseries 130.55&163.21&135.75&137.25&147.43&3201.22&4189.81&634.84&1609.07\\
gnm-undir'-16777216&\textsc{Lex}&59596.38&\em OOT&\em OOT&66.52&53.63&68.61&61.56&\bfseries 50.05&66.84&1221.36&1367.83&191.17&479.07\\
gnm-undir'-16777216&\textsc{Top50}&59678.65&\em OOT&\em OOT&326.51&154.37&167.29&164.44&\bfseries 148.69&173.85&3245.67&3316.86&668.81&991.27\\
gnm-undir'-18874368&\textsc{Out50}&81000.73&\em OOT&\em OOT&231.06&\bfseries 149.48&171.70&165.07&152.66&176.98&3599.90&4274.51&737.59&1756.56\\
gnm-undir'-18874368&\textsc{Lex}&81551.57&\em OOT&\em OOT&73.84&61.92&74.01&64.26&\bfseries 58.64&72.29&1370.50&1575.95&213.04&566.16\\
gnm-undir'-18874368&\textsc{Top50}&80884.17&\em OOT&\em OOT&390.55&\bfseries 169.78&194.55&176.25&173.54&188.13&3569.38&3831.89&755.08&1206.42\\
gnm-undir'-25165824&\textsc{Out50}&\em OOT&\em OOT&\em OOT&363.84&210.36&242.81&\bfseries 207.91&210.87&242.69&4749.63&6355.28&995.79&2332.60\\
gnm-undir'-25165824&\textsc{Lex}&\em OOT&\em OOT&\em OOT&97.38&84.32&110.47&92.75&\bfseries 83.75&110.92&1754.68&2212.26&301.51&741.83\\
gnm-undir'-25165824&\textsc{Top50}&\em OOT&\em OOT&\em OOT&610.22&\bfseries 239.12&275.35&244.99&244.97&258.77&4886.79&5197.86&1149.16&1532.49\\
gnm-undir'-28311552&\textsc{Out50}&\em OOT&\em OOT&\em OOT&436.47&\bfseries 239.34&286.98&257.29&240.93&285.76&5361.47&7146.88&1100.41&2687.84\\
gnm-undir'-28311552&\textsc{Lex}&\em OOT&\em OOT&\em OOT&116.75&95.96&121.04&94.05&\bfseries 92.68&127.10&1978.30&2458.86&322.52&998.63\\
gnm-undir'-28311552&\textsc{Top50}&\em OOT&\em OOT&\em OOT&750.54&296.07&330.78&\bfseries 284.67&287.72&313.37&5577.72&5865.60&1162.99&1790.99\\
com-Orkut&\textsc{Out50}&6937.69&\em OOT&\em OOT&\bfseries 1254.98&1288.93&1888.63&1330.74&1301.37&1830.77&27374.58&37763.69&7666.46&17267.06\\
com-Orkut&\textsc{Lex}&6840.47&3430.28&4282.09&\bfseries 329.49&330.00&683.75&335.90&336.43&654.91&3073.72&10525.33&928.15&7714.35\\
com-Orkut&\textsc{Top50}&6662.29&34305.13&\em OOT&401.25&\bfseries 386.90&744.37&405.66&395.34&766.86&4623.05&12880.96&1305.60&8691.12\\
hollywood-2009&\textsc{Out50}&3491.90&36865.03&\em OOT&642.46&559.26&789.07&577.09&\bfseries 553.47&796.91&14392.87&21300.98&3007.11&9752.45\\
hollywood-2009&\textsc{Lex}&3468.54&1436.24&3419.62&181.91&154.65&300.20&156.75&\bfseries 149.74&317.31&2437.27&7978.86&499.69&4868.87\\
hollywood-2009&\textsc{Top50}&3451.53&6538.26&10702.41&215.24&\bfseries 175.03&312.50&179.51&180.06&315.37&2950.63&7833.03&620.79&5023.17\\
mycielskian17&\textsc{Out50}&64370.09&\em OOT&\em OOT&558.08&\bfseries 407.29&574.99&467.34&430.88&652.29&12311.93&21074.98&2653.99&7227.03\\
mycielskian17&\textsc{Lex}&63497.86&\em OOT&\em OOT&149.09&132.13&227.70&132.28&\bfseries 126.58&226.57&3018.01&8649.85&526.16&1833.20\\
mycielskian17&\textsc{Top50}&64031.41&\em OOT&\em OOT&455.45&\bfseries 305.41&438.83&315.91&311.31&418.88&7649.29&12719.26&1679.07&2957.50\\\bottomrule
    \end{tabular}}
    \caption{Running time on all instances over three repetitions for the competitors by Borowitz~et~al.~\cite{dblp:conf/acda/borowitzg023} (\textsf{BFS20}) and Großmann~et~al.~\cite{grossmann2025engineering} (\textsf{Im(provedDynOpt)},\textsf{Strong(DynOpt)}) and our implementations of the $\lambda$-fair algorithm by Chekuri~et~al.~\cite{chekuri2024adaptive}.
    F=\textsf{Fractional},P=\textsf{PackedFractional} and PL=F=\textsf{PackedFractional, list}. \emph{OOT} means that the computation did not finish in less than 24 hours.}
    \label{tab:results:runtime}
\end{sidewaystable}
\clearpage
\section{Instance Details}
\label{app:features:graph}
\begin{table}[h]
    \centering
        \caption{Detailed per-instance properties of the dense set.}
    \label{tab:graph_properties}
    \centering
    \begin{tabular}{llrrrrr}\toprule
    &&&&\multicolumn{2}{c}{Degree}\\\cmidrule(lr){5-6}
  \bf Instance& \bf Mode  & $m$& $n$ & $\max$ & $\min$  & \bf  Components\\
  \midrule
gnm-undir'-16777216& \textsc{Lex} & \numprint{16777216} & \numprint{8192} & \numprint{4257} & \numprint{3923} & \numprint{1}\\
gnm-undir'-16777216& \textsc{Lex+Out50} & \numprint{8388608} & \numprint{8192} & \numprint{4137} & \numprint{5} & \numprint{1}\\
gnm-undir'-16777216& \textsc{Lex+Top50} & \numprint{8388608} & \numprint{8192} & \numprint{3190} & \numprint{1030} & \numprint{1}\\
gnm-undir'-18874368& \textsc{Lex} & \numprint{18874368} & \numprint{8192} & \numprint{4766} & \numprint{4458} & \numprint{1}\\
gnm-undir'-18874368& \textsc{Lex+Out50} & \numprint{9437184} & \numprint{8192} & \numprint{4638} & \numprint{7} & \numprint{1}\\
gnm-undir'-18874368& \textsc{Lex+Top50} & \numprint{9437184} & \numprint{8192} & \numprint{3591} & \numprint{1158} & \numprint{1}\\
gnm-undir'-25165824& \textsc{Lex} & \numprint{25165824} & \numprint{8192} & \numprint{6289} & \numprint{6006} & \numprint{1}\\
gnm-undir'-25165824& \textsc{Lex+Out50} & \numprint{12582912} & \numprint{8192} & \numprint{6149} & \numprint{3} & \numprint{1}\\
gnm-undir'-25165824& \textsc{Lex+Top50} & \numprint{12582912} & \numprint{8192} & \numprint{4731} & \numprint{1539} & \numprint{1}\\
gnm-undir'-28311552& \textsc{Lex} & \numprint{28311552} & \numprint{8192} & \numprint{7054} & \numprint{6781} & \numprint{1}\\
gnm-undir'-28311552& \textsc{Lex+Out50} & \numprint{14155776} & \numprint{8192} & \numprint{6926} & \numprint{4} & \numprint{1}\\
gnm-undir'-28311552& \textsc{Lex+Top50} & \numprint{14155776} & \numprint{8192} & \numprint{5278} & \numprint{1731} & \numprint{1}\\
com-Orkut& \textsc{Lex} & \numprint{117185083} & \numprint{3072441} & \numprint{33313} & \numprint{1} & \numprint{1}\\
com-Orkut& \textsc{Lex+Out50} & \numprint{25537869} & \numprint{3072441} & \numprint{10978} & \numprint{0} & \numprint{50844}\\
com-Orkut& \textsc{Lex+Top50} & \numprint{111022030} & \numprint{3072441} & \numprint{24704} & \numprint{1} & \numprint{1}\\
hollywood-2009& \textsc{Lex} & \numprint{56375711} & \numprint{1139905} & \numprint{11467} & \numprint{0} & \numprint{44508}\\
hollywood-2009& \textsc{Lex+Out50} & \numprint{4938745} & \numprint{1139905} & \numprint{636} & \numprint{0} & \numprint{322934}\\
hollywood-2009& \textsc{Lex+Top50} & \numprint{54474408} & \numprint{1139905} & \numprint{10994} & \numprint{0} & \numprint{44508}\\
mycielskian17& \textsc{Lex} & \numprint{50122871} & \numprint{98303} & \numprint{49151} & \numprint{16} & \numprint{1}\\
mycielskian17& \textsc{Lex+Out50} & \numprint{12142982} & \numprint{98303} & \numprint{30491} & \numprint{0} & \numprint{2104}\\
mycielskian17& \textsc{Lex+Top50} & \numprint{32748134} & \numprint{98303} & \numprint{42130} & \numprint{16} & \numprint{1}\\\bottomrule
    \end{tabular}

\end{table}

\end{document}